# Electronic band structures and intra-atomic interactions in layered quaternary oxyarsenides LaZnAsO and YZnAsO

V.V. Bannikov, I.R. Shein, A.L. Ivanovskii *

*Institute of Solid State Chemistry, Ural Branch of the Russian Academy of Sciences, 620041, Ekaterinburg, Russia*

**Abstract**

First-principle FLAPW-GGA band structure calculations are employed to obtain the structural, electronic properties and chemical bonding picture for two related layered phases, namely, quaternary oxyarsenides LaZnAsO and YZnAsO. These compounds are found to be direct-transition type semiconductors with the GGA gaps of about 0.65-1.30 eV. The peculiarities of chemical bonding in these phases are investigated and discussed in comparison with quaternary oxyarsenide LaFeAsO - a basic phase for the newly discovered 26-52K superconductors.

* Corresponding author.
*E-mail address:* shein@ihim.uran.ru (I.R. Shein).

# 1. Introduction

The recent discovery [1-6] of a new group of the so-called FeAs-superconductors with transition temperatures to $T_C \sim 55K$ based on layered quaternary oxypnictides (*LnFePn*O, where *Ln* are early rare earth metals such as La, Ce, Sm, Dy, Gd, and *Pn* are P or As) has sparked enormous interest in this group of materials.

Besides superconductivity, these materials, which belong to a much larger family of quaternary systems with the tetragonal ZrCuSiAs type structure (space group *P4/nmm*) [7], possess broadly varying properties depending on their chemical composition - such as magnetic ordering, optical and opto-electronic characteristics *etc*, see review [7].

The ZrCuSiAs-like oxypnictides *LnMPn*O (where *M* are transition metals) have a quasi-two-dimensional crystal structure with alternating [*MPn*] and [*Ln*O] molecular layers, where the *M* atoms are arranged on a simple square lattice [1,7]. Previous studies [7-12] indicate that the electronic bands in LaFeAsO and related phases in the window around the Fermi level are formed mainly by the states of [Fe*Pn*] layers, while the [*Ln*O] layers provide the charge reservoir when doped with various ions.

Thus the properties of quaternary oxypnictides *LnMPn*O will change drastically if the type of the transition metal *M* is varied. For example, depending on the type of the *M* sublattice, the oxyphosphides La*M*PO behave as semiconductors (for Mn), superconductors (for Fe) or metals (for Co). Among the oxyarsenides La*M*AsO, there are magnetic (for *M* = V, Cr, Mn and Co) and non-magnetic (for *M* = Ni and Cu) systems. In turn, among magnetic materials there are an antiferromagnetic semiconductor (LaMnAsO) and a ferromagnetic metal (LaCoAsO) [13].

Quite recently, nine new Zn-containing oxyarsenides, namely *Ln*ZnAsO (*Ln* = Y, La-Nd, Sm, Gd-Dy) have been synthesized as single crystals [14] by the reaction of *Ln*, ZnO and arsenic in a NaCl/KCl flux, and their crystal structure (of the tetragonal ZrCuSiAs-like type) was determined. It was proposed that these species are metallic-like [14]. On the other hand, carrier transport experiments and optical absorption spectra of LaZnAsO (prepared as films on (001) MgO substrates by reactive solid-phase epitaxy) exhibit a semiconducting - like behavior of this compound [15]; it is confirmed also by DFT calculations for the isoelectronic LaZnAsP [16]. It is expected also [15] that LaZnAsO may be a promising matrix for diluted magnetic semiconductors if $Zn^{2+}$ is partially replaced by a divalent transition metal, for example $Mn^{2+}$.

Besides, the peculiarities of chemical bonding in zinc - containing oxyarsenides - in comparison with *LnM*AsO (where *M* are transition metals with partially filled $d^n$ shells) - should be rather interesting. Really, for quaternary *LnM*AsO a very complex picture of chemical binding takes place, where along with ionic inter-layer interactions (due to electron transfer $[LnO]^{\delta+} \rightarrow [MAs]^{\delta-}$) which are responsible for the stabilization of these compounds, the mixed metallic-ionic-covalent intra-layer bonds with the participation of valence states of *M* atoms occur [7,17,18].

In the present study, using the first principles FLAPW method within the generalized gradient approximation (GGA) for the exchange-correlation potential we explore for the first time the electronic properties for two quaternary oxyarsenides LaZnAsO and YZnAsO. In addition, the peculiarities of the chemical bonding for Zn-containing systems are investigated and discussed in comparison with those for LaFeAsO as a parent phase for the newly discovered superconductors.

## 2. Computational details

Our calculations were carried out by means of the full-potential method with mixed basis APW+lo (FLAPW) implemented in the WIEN2k suite of programs [19]. The generalized gradient approximation (GGA) to exchange-correlation potential in the PBE form [20] was used. The plane-wave expansion was taken up to $R_{MT} \times K_{MAX}$ equal to 7, and the *k* sampling with 13×13×5 *k*- points in the Brillouin zone was used. La ($4f^0 5s^2 5p^6 5d^1 6s^2$), Y ($4s^2 4p^6 4d^1 5s^2$), Zn ($3d^{10} 4s^2 4p^0$), O ($2s^2 2p^4$) and As ($4p^3 4s^2$) were treated as valence states.

The calculations were performed with full-lattice optimizations including the so-called internal parameters $z_{La(Y)}$ and $z_{As}$, see [7]. The self-consistent calculations were considered to be converged when the difference in the total energy of the crystal did not exceed 0.1 mRy and the difference in the total electronic charge did not exceed 0.001 $e$ as calculated at consecutive steps.

The analysis of the hybridization effects was performed using the densities of states (DOS), which were obtained by a modified tetrahedron method [21]. To quantify the amount of electrons redistributed between the atomic sublattices and the adjacent layers, *i.e.* for the discussion of the ionic bonding, we performed also a Bader [21] analysis.

## 3. Results and discussion

*3.1. Structural properties.*

As the first step, the total energy ($E_{tot}$) *versus* cell volume calculations were carried out to determine the equilibrium structural parameters for the examined LaZnAsO and YZnAsO; the calculated values are presented in Table 1. These data are in reasonable agreement with the available experiment [13]: the divergences $(a^{calc} - a^{exp})/a^{exp}$ and $(c^{calc} - c^{exp})/c^{exp}$ for LaZnAsO and YZnAsO are 0.0002, 0.0053 and 0.0090, 0.0093, respectively. These divergences between the calculated and experimentally observed parameters should be attributed to the well known overestimation of structural parameters for GGA calculations.

As can be seen, the parameters *a* and *c* for YZnAsO are smaller than those for LaZnAsO, *i.e.* when going from LaZnAsO to YZnAsO the inter-layer distances decrease and simultaneous compression of the layers in the *xy* planes takes place. This result can be easily explained taking into account the circumstance that the replacement of Y (atomic radius $R^{at}$ = 1.81 Å) by larger La atoms ($R^{at}$ = 1.87 Å) results in stretching of the corresponding chemical bonds. Nevertheless, some *anisotropic deformation* of the crystal structure takes place as going from LaZnAsO to YZnAsO: according to our calculations, the change of the parameter *c* (about 2.5%) is smaller than for the parameter *a* (about 3.2%). The same tendency is observed in the experiment, see Table 1.

*3.2. Electronic structure.*

Figure 1 shows the band structures of LaZnAsO and YZnAsO as calculated along the high-symmetry *k* lines. The $E(k)$ curves in the high-symmetry directions in the Brillouin zone (BZ) demonstrate evident similarities in the energy bands for these isoelectronic and isostructural oxyarsenides. However the bands dispersion in YZnAsO is stronger than in LaZnAsO: the valence band (VB) width in YZnAsO is about 0.7 eV larger. Another obvious difference is that the band gap (BG) for YZnAsO (1.3 eV) is twice as large as for LaZnAsO (0.65 eV). Note that the VB maximum and the conduction band minimum are located at the Γ point ($k = 0$) indicating that both systems are direct-transition type semiconductors.

At the same time, the experimentally obtained direct-transition type BG for LaZnAsO thin film is observed at about 1.5 eV [14]. This divergence is related to the well-known underestimation of the BG values within LDA-GGA based calculation methods. A standard empirical correction requires fitting of the LDA-GGA gap to experimental values, see for example [22]. In our case, for LaZnAsO a multiplicative correction factor (2.14) approximates the calculated BG to the experimentally measured gap [14]. Using this correction factor, the "experimental" gap for YZnAsO was estimated from our calculated data to be 2.78 eV.

Figure 2 shows the total and atomic-resolved *l*-projected DOSs for LaZnAsO and YZnAsO as calculated for equilibrium geometries. For both oxypnictides the quasi-core As 4*s* states are placed in range from - 12 to -10 eV below the Fermi level. The fully occupied Zn 3*d* states are located from - 8 to -6.5 eV and are separated from the near-Fermi valence band by a gap. In turn, the valence states occupy energy intervals from the Fermi level down to -5.4 and -5.2 eV in LaZnAsO and YZnAsO, respectively, and in this region the admixtures of contributions from all atoms take place. Note that La $f$ states form an intensive unoccupied DOS peak near the bottom of the conduction band of LaZnAsO.

Thus, the preliminary conclusion from our DOSs calculations is that the general bonding mechanism in both oxyarsenides is not of a "purely" ionic character, but owing to the mentioned hybridization of valence states includes also covalent interactions. Let us discuss the bonding picture for LaZnAsO and YZnAsO in greater detail.

*3.3. Chemical bonding.*

**The covalent bonding** character of LaZnAsO and YZnAsO phases may be well understood from site-projected DOS calculations. As is shown in Fig. 2, Zn-As and La(Y)-O states are hybridized. In addition, taking into account the inter-atomic distances in these phases, As-As bonds should be assumed to be covalent, see also [18]. These covalent bonds are clearly visible in Fig. 3, where the charge density maps for LaZnAsO and YZnAsO are depicted in comparison with LaFeAsO. The charge density distribution in the (110) planes reveals the mentioned bonding between As atoms (inside [ZnAs] layers) and additional low electron density between (La,Y) and As atoms. This implies that weak inter-layer covalent La(Y)-As bonds occur. The charge density distribution in the (400) planes demonstrates the formation of Zn-As bonds (inside [ZnAs] layers), which weaken as going from YZnAsO to LaZnAsO. This result can be easily explained taking into account the growth of the inter-atomic distances for LaZnAsO as compared with YZnAsO, Table 1.

To describe the **ionic bonding** for LaZnAsO and YZnAsO, it is possible to begin with a simple ionic picture, which considers the usual oxidation numbers of atoms: $(La,Y)^{3+}$, $Zn^{2+}$, $As^{3-}$ and $O^{2-}$. Thus, the charge states of the layers are $[La(Y)O]^{1+}$ and $[ZnAs]^{1-}$, *i.e.* the charge transfer occurs from [La(Y)O] to [ZnAs] layers. Besides, inside [La(Y)O] and [ZnAs] layers, the ionic bonding takes place respectively between La(Y)-O and Zn-As.

To estimate the amount of electrons redistributed between the adjacent [La(Y)O] and [ZnAs] layers and between the atoms inside each layer, we carried out a Bader [22] analysis. In this approach, each atom of a crystal is surrounded by an effective surface that runs through minima of the charge density, and the total charge of an atom (the so-called Bader charge, $Q^B$) is determined by integration within this region. The calculated atomic $Q^B$ as well as the corresponding charges as obtained from the purely ionic model ($Q^i$) and their differences ($\Delta Q = Q^B - Q^i$) are presented in Table 2. These results show that the inter-layer charge transfer is much smaller than it is predicted in the idealized ionic model. Namely, the transfer from [LaO] to [ZnAs] layers is about 0.36 and from [YO] to [ZnAs] layers – about 0.31 electrons per formula unit for LaZnAsO and YZnAsO respectively. For LaFeAsO, this transfer is higher: 0.39 electrons per a formula unit.

Thus, summarizing the above results, the picture of chemical bonding for LaZnAsO and YZnAsO may be described as follows.

(i) Inside [La(Y)O] layers, mixed ionic-covalent bonds La(Y)-O take place (owing to hybridization of valence states of La(Y) and oxygen atoms and La(Y) → O charge transfer);

(ii) Inside [ZnAs] layers, mixed ionic-covalent bonds Zn-As appear (owing to hybridization of valence states of Zn and As atoms and Zn → As charge transfer);

(iii) In addition, inside [ZnAs] layers pure covalent bonds As-As take place (owing to As 4*p* - As 4*p* hybridization);

(iv) Between the adjacent [La(Y)O] and [ZnAs] layers, ionic bonds emerge owing to [La(Y)O] → [ZnAs] charge transfer and weak covalent "intra-layer" La(Y)-As bonds take place owing to hybridization of valence states of La(Y) and As atoms.

(v) Generally, the bonding in LaZnAsO and YZnAsO can be classified as a mixture of ionic and covalent contributions.

Note that the main difference in the bonding pictures for LaZnAsO, YZnAsO *versus* LaFeAsO consists in intra-atomic interactions in [ZnAs] *versus* [FeAs] layers. Really, as distinct from LaZnAsO and YZnAsO, mixed metallic-ionic-covalent bonds occur in LaFeAsO inside [FeAs] layers owing to participation of Fe atoms in Fe-As covalent bonds (as a result of Fe $3d$ - As $4p$ hybridization), Fe-Fe metallic-like bonds (due to delocalized Fe $3d$ states), and in Fe-As ionic bonds (due to Fe → As charge transfer), see also [7,16,17].

## 4. Conclusions

In summary, we have performed FLAPW-GGA calculations to obtain the structural, electronic properties and the chemical bonding picture for two quaternary oxyarsenides - LaZnAsO and YZnAsO. Both materials are found to be direct-transition type semiconductors with the GGA band gaps at about 0.65 eV for LaZnAsO and 1.3 eV for YZnAsO.

Our results indicate also that the bonding in the Zn-containing oxyarsenides LaZnAsO and YZnAsO can be classified as a mixture of ionic and covalent contributions - unlike for LaFeAsO (and other oxypnictides with participation of transition metals with partially filled $d^n$ shells), where mixed metallic-ionic-covalent bonds occur.

**References**

[1] Y. Kamihara, T. Watanabe, M. Hirano, H. Hosono, J. Am. Chem. Soc. 130 (2008) 3296.
[2] H. Takahashi, K. Igawa, K. Arii, Y. Kamihara, M. Hirano, H. Hosono, Nature 453 (2008) 376.
[3] Z.A. Ren, J. Yang, W. Lu, W. Yi, X.L. Shen, Z.C. Li, G.C. Che, X.L. Dong, L.L. Sun, F. Zhou, Z.H. Zhao, Euro. Phys. Lett. 82 (2008) 57002.
[4] X.H. Chen, T. Wu, G. Wu, R.H. Liu, H. Chen, D. Fang, Nature 453 (2008) 761.
[5] C. Cruz, Q. Huang, J.W. Lynn, J. Li, W. Ratcli, J.L. Zarestky, H.A. Mook, G.F. Chen, K.L. Luo, N.L. Wang, P. Dai, Nature 453 (2008) 899.
[6] A. Cho, Science 320 (2008) 433.
[7] D. Johrendt, R. Pottgen, Angew. Chem. Int. Ed. 47 (2008) 4782.
[8] S. Lebègue, Phys. Rev. B75 (2007) 2007.
[9] D.J. Singh, M.H. Du, Phys. Rev. Lett. 100 (2008) 237003.
[10] I.R. Shein, A.L. Ivanovskii, Phys. Rev. B78 (2008) 104519.
[11] Z.P.Yin, S. Lebègue, M. J. Han, B. Neal, S.Y. Savrasov, W.E. Pickett, Phys. Rev. Lett. 101(2008) 047001.
[12] I.R. Shein, A.L. Ivanovskii, JETP Lett. 88 (2008) 115.
[13] G. Xu, W. Ming, Y. Yao, X. Dai, S.C. Zhang, Z. Fang, arXiv:0803.1282 (2008).
[14] A.T. Nientiedt, W. Jeitschko, Inorganic Chem. 37 (1998) 386.
[15] K. Kayanuma, R. Kawamura, H. Hiramatsu, H. Yanagi, M. Hirano, T. Kamiya, H. Hosono, Thin Solid Films 516 (2008) 5800.
[16] K. Kayanuma, H. Hiramatsu, M. Hirano, R. Kawamura, H. Yanagi, T. Kamiya, H. Hosono, Phys. Rev. B76 (2007) 195325.
[17] I.R. Shein, A.L. Ivanovskii, Scripta Materialia 59 (2008) 1099.
[18] T. Yildirim, arXiv:0807.3936 (2008).
[19] P. Blaha, K. Schwarz, G. Madsen et al. WIEN2k, *An Augmented Plane Wave Plus Local Orbitals Program for Calculating Crystal Properties, Vienna University of Technology,* Vienna, 2001.
[20] J.P. Perdew, S. Burke, M. Ernzerhof, Phys. Rev. Lett. 77 (1996) 3865.
[21] P.E. Blochl, O. Jepsen, O.K. Anderson, Phys. Rev. B 49 (1994) 16223.
[22] R.F.W. Bader, *Atoms in Molecules: A Quantum Theory, International Series of Monographs on Chemistry,* Clarendon Press, Oxford, 1990.
[22] J. Robertson, K. Xiong, S.J. Clark, Thin Solid Films 496 (2006) 1.

Table 1.
The optimized lattice parameters (*a* and *c*, in Å) and internal coordinates ($z_{La(y)}$ and $z_{As}$) for LaZnAsO and YZnAsO.

| phase/parameter | LaZnAsO | YZnAsO |
|---|---|---|
| *a* | 4.0908 (4.090 [13]) | 3.9600 (3.943 [13]) |
| *c* | 9.1495 (9.068 [13]) | 8.9250 (8.843 [13]) |
| *c/a* | 2.2366 (2.2144 [13]) | 2.2626 (2.2427 [13]) |
| $z_{La}$ | 0.1341 | 0.1215 |
| $z_{As}$ | 0.6698 | 0.6816 |

* available experimental data are given in parentheses.

Table 2.
Atomic charges (in e) for LaZnAsO and YZnAsO and for [ZnAs], and [La(Y)O] layers as obtained from a purely ionic model ($Q^i$), Bader analysis ($Q^B$) and their differences ($\Delta Q = Q^B - Q^i$).

|  |  | La(Y) | Zn(Fe) | As | O | [La(Y)O] | [Zn(Fe)As] |
|---|---|---|---|---|---|---|---|
| LaZnAsO | $Q^i$ | +3 | +2 | -3 | -2 |  |  |
|  | $Q^B$ | 9.069 | 11.532 | 6.110 | 7.289 | 16.358 | 18.821 |
|  | $\Delta Q$ | 1.069 | 1.532 | -1.890 | -0.771 | 0.358 | -0.358 |
| YZnAsO | $Q^i$ | +3 | +2 | -3 | -2 |  |  |
|  | $Q^B$ | 8.938 | 11.550 | 6.136 | 7.376 | 16.314 | 18.928 |
|  | $\Delta Q$ | 0.938 | 1.550 | -1.864 | -0.624 | 0.314 | -0.314 |
| LaFeAsO | $Q^i$ | +3 | +2 | -3 | -2 |  |  |
|  | $Q^B$ | 9.115 | 7.722 | 5.887 | 7.276 | 16.391 | 14.998 |
|  | $\Delta Q$ | 1.115 | 1.772 | -2.113 | -0.724 | 0.391 | -0.391 |

FIGURES

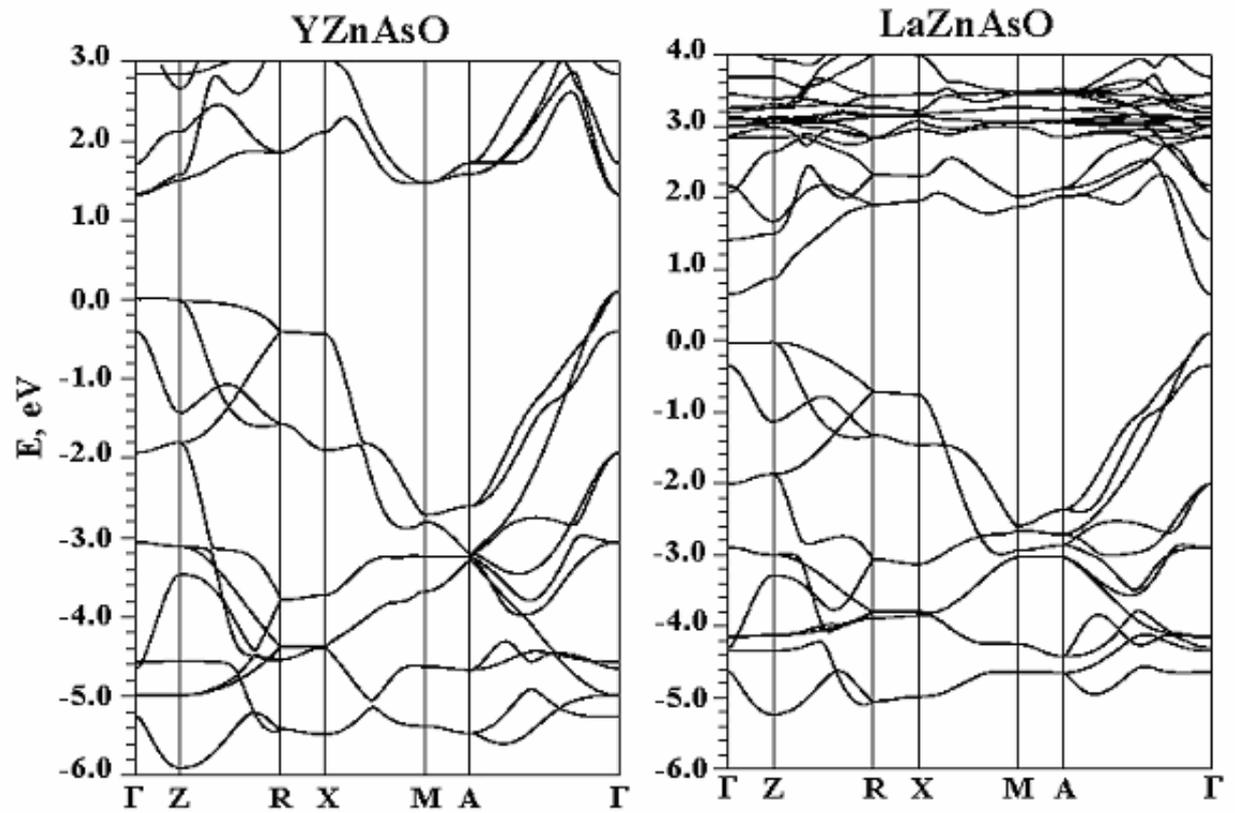

**Figure 1.** Electronic bands for LaZnAsO and YZnAsO.

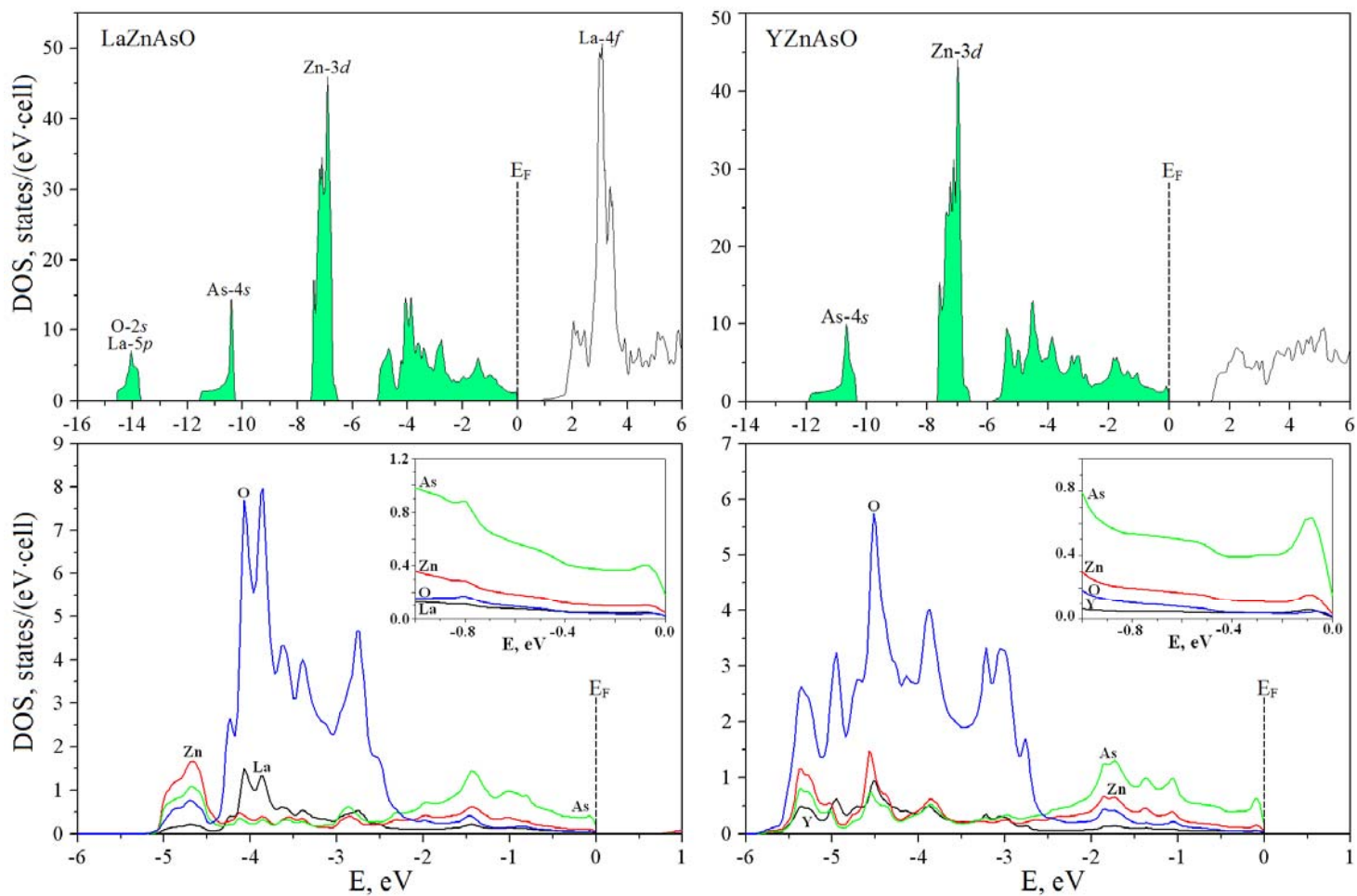

**Figure 2**. (*Color online*) Total (*upper panels*) and partial densities of states (*bottom panels*) for LaZnAsO and YZnAsO

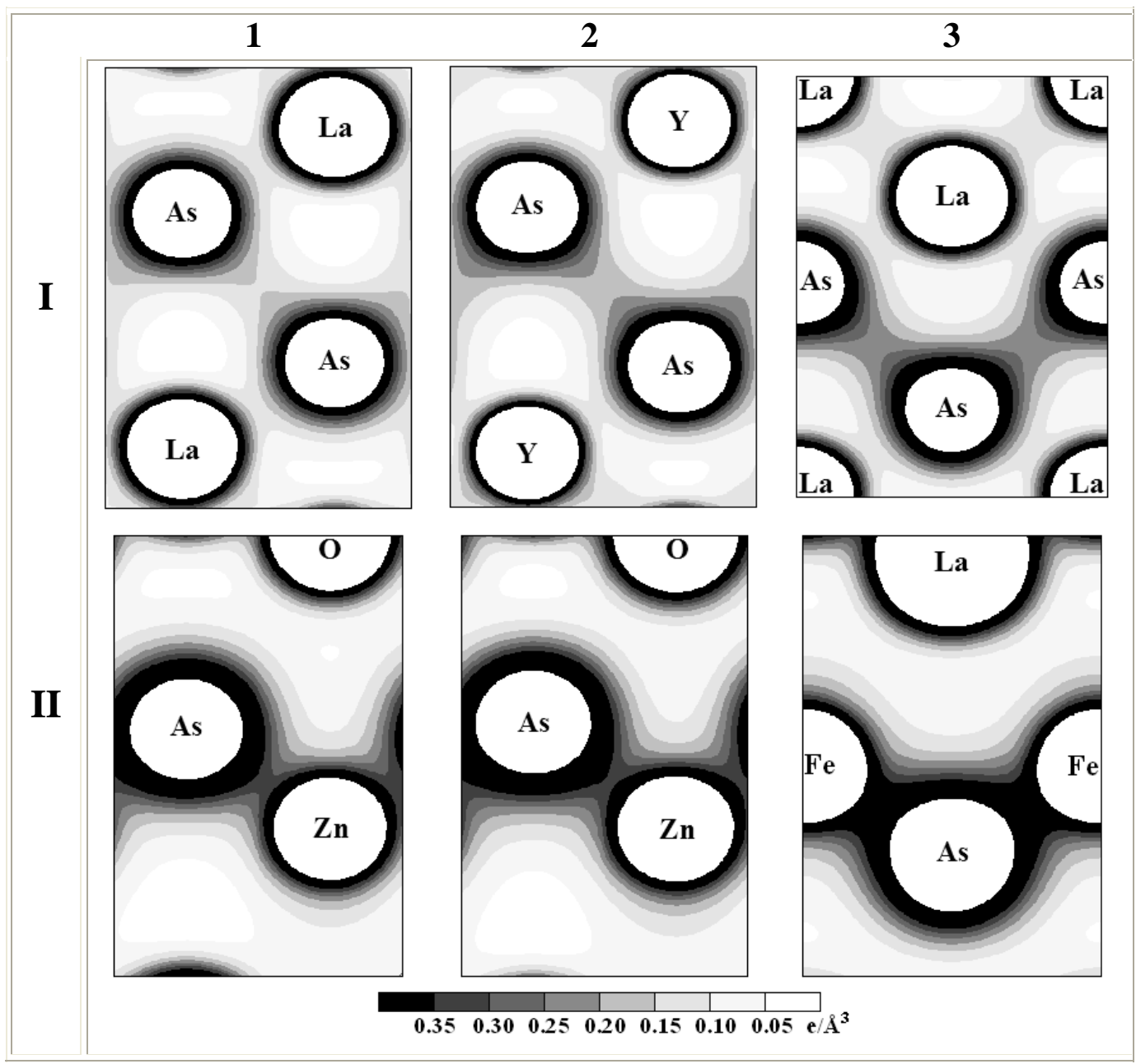

**Figure 3.** Electron density distribution in LaZnAsO (1) and YZmAsO (2) for the (110) and (400) planes as compared with LaFeAsO (3) for the (100) and (200) planes - panels I and II, respectively.

13